\documentclass [aps,prb,superscriptaddress,twocolumn,showpacs,
a4paper,unsortedaddress,amsmath,amssymb,byrevtex,floatfix]{revtex4}

\usepackage[latin1]{inputenc}
\usepackage{graphicx}
\usepackage{dcolumn}
\usepackage{bm}
\usepackage{hyperref}
\usepackage{times}

\begin{document}

\title{Impact of Fano and Breit-Wigner resonances in the thermoelectric
properties of nanoscale junctions}

\author{V. M. Garc\'{\i}a-Su\'arez}
\affiliation{Departamento de F\'{\i}sica, Universidad de Oviedo,
33007 Oviedo, Spain}
\affiliation{Nanomaterials and Nanotechnology Research Center (CINN),
Oviedo, Spain}
\affiliation{Department of Physics, Lancaster University, Lancaster,
LA1 4YB, United Kingdom}

\author{R. Ferrad\'as}
\affiliation{Departamento de F\'{\i}sica, Universidad de Oviedo,
33007 Oviedo, Spain}
\affiliation{Nanomaterials and Nanotechnology Research Center (CINN),
Oviedo, Spain}

\author{J. Ferrer}
\affiliation{Departamento de F\'{\i}sica, Universidad de Oviedo,
33007 Oviedo, Spain}
\affiliation{Nanomaterials and Nanotechnology Research Center (CINN),
Oviedo, Spain}
\affiliation{Department of Physics, Lancaster University, Lancaster,
LA1 4YB, United Kingdom}

\date{\today}

\begin{abstract}
We show that the thermoelectric properties of nanoscale junctions featuring
states near the Fermi level strongly depend on the
type of resonance generated by such states, which can be either Fano or
Breit-Wigner-like. We give general expressions for the
thermoelectric coefficients generated by the two types of resonances and
calculate the thermoelectric properties of these systems,
which encompass most nanoelectronics junctions. We include simulations of
real junctions where metalloporphyrin dithiolate molecules bridge gold electrodes
and prove that for some metallic elements the thermoelectric properties show a
large variability with respect to the position of the resonance near the Fermi
level. We find that the thermopower and figure of merit are largely
enhanced when the resonance gets close to the Fermi level and reach values
higher than typical values found in other nanoscale junctions.
The specific value and temperature dependence are determined by a series
of factors such as the strength of the coupling between the state and
other molecular states, the symmetry of the state, the strength of the
coupling between the molecule and the leads and the spin filtering
behavior of the junction.
\end{abstract}

\pacs{72.20.Pa,73.63.-b,85.35.Ds,71.15.-m}

\maketitle

\section{Introduction}

The use of molecules as electronic components is expected to surpass,
at least temporally, one of the limits imposed to Moore's law as the size
of the electronic components shrinks towards the atomic limit.
Their particular properties would allow to develop more involved and
efficient circuits and electronic devices with sizes much smaller that
conventional silicon-based devices\cite{Fer09}. Among the properties
that molecules in metallic junctions can show are
rectification\cite{Avi74}, negative differential resistance\cite{Gar08},
switching\cite{Piv05}, memory\cite{He06} and sensing\cite{Lon06}.
Adding a thermal gradient or a coupling to a thermal bath\cite{Ent10,Jor13}
would also allow molecules to work as nanometer-size thermoelectric
devices\cite{Dub11}, which could be used in applications ranging from
chip cooling to building refrigerators. One of the thermoelectric
coefficients, the Seebeck coefficient $S$, is also specially suited to
gather information on the mechanisms of molecular conduction\cite{Koc04}
and the chemistry of the junction\cite{Bah08}. For instance, from the
sign of $S$ it is possible to deduce if the Fermi level lies close to
the HOMO or the LUMO orbital\cite{Pau03,Tan10}. A positive (negative)
sign indicates $p$-type ($n$-type) conduction, which means the Fermi
level lies near the HOMO (LUMO). This implies that the sign and
magnitude of the thermopower can be changed by gating the
molecule\cite{Zhe04,Wan05,Wie10}. The Seebeck coefficient is also a very
sensible magnitude that depends on factors such as the molecular
length\cite{Pau08,Tan11}, the molecular conformation\cite{Bur12}, the
contact group\cite{Bal12}, the side groups\cite{Fin08,Liu11,Sta11}, the
surface reconstruction\cite{Hsu12} and the type of electrodes\cite{Noz10}.

The most important quantity that measures the thermoelectric efficiency
of a system is the dimensionless figure of merit $ZT=S^2GT/\kappa$,
which is proportional to the square of the Seebeck coefficient ($S$)
and the conductance ($G$), and inversely proportional to the thermal
conductance ($\kappa$). This number, which determines how easy it is
to transform heat into electricity, should be as high as possible
(closer to 1 or higher) in order for a thermoelectric device to work
effectively. Values larger than 1 ($\sim 2.4$) have already been
measured in inorganic superlattice devices\cite{Ven01}. In the field of
molecular electronics, however, despite current
efforts\cite{Red07,Bah08,Mal10,Yee11,Bub11,Wid11}, the measured $S$
and $ZT$ are not yet very high ($|S|\sim 33$ $\mu$V/K\cite{Yee11} and
$ZT\sim 0.25$\cite{Bub11}). Theoretical calculations predict that
much higher values should be achieved when Fano resonances\cite{Fin08,Tro12}
or interference-related peaks\cite{Kar11} cross the Fermi level, but
such predictions have not been confirmed experimentally yet. These
values, calculated in the framework of coherent transport, should
also be corrected by taking into account the phonon thermal
conductance\cite{Mur08}, inelastic scattering\cite{Pop12} and, in
general, coupling to phonons\cite{Lei10}.

In this article we calculate the thermoelectric coefficients and
figure of merit of molecules which show intereference-related features
around the Fermi level\cite{Kal12} in the form of Breit-Wigner-like and
Fano resonances. In section \ref{ThCoeff} we give a brief theoretical
introduction on the thermoelectric coefficients. In next section,
\ref{FirstP}, we include ab-initio simulations of molecular wires
(metalloporphyrins dithiolate). Finally, in section \ref{TightB} we present a
model that can be used to study the evolution of the thermoelectric properties
of molecular wires as a function of a series of parameters, among which
are included the type of resonance that crosses the Fermi level, which is
related to the symmetry of the state, the strength of coupling between
the resonance and other molecular levels and the coupling between the
molecule and the electrodes.

\section{Thermoelectric coefficients}\label{ThCoeff}

When a junction is subject to an electrostatic potential difference
and a temperature gradient, electric and heat current flow from one
electrode to another. With just an electrostatic potential, electrons
move from the negative to the positive electrode, whereas the electric
current is defined to flow in the opposite direction. With a
temperature gradient, however, there is no rule of thumb that allows
one to determine how the current flows without accurate information on
the electronic structure, i.e. the transmission, of the junction. If
the transmission below the Fermi level is higher, electrons flow from
the cold to the hot electrode, whereas the contrary happens if the
transmission is higher above the Fermi level. The electrostatic
potential and the temperature gradient generate also a heat
current which in general flows from the hot to the cold electrode.
For a system with spin polarization\cite{Mas10,SpinPol}, this is
summarized in the following equation:

\begin{equation}\label{Eq1}
\left(\begin{array}{c}I\\\dot Q\end{array}\right)=\frac{1}{h}
\left(\begin{array}{cc}e^2L_0^\mathrm{t}&
\frac{e}{T}L_1^\mathrm{t}\\
eL_1^\mathrm{t}&\frac{1}{T}L_2^\mathrm{t}\end{array}\right)
\left(\begin{array}{c}\Delta V\\\Delta T
\end{array}\right)
\end{equation}

\noindent where $I$ and $\dot Q$ are the electric and heat currents,
respectively, and the moments $L_n^\mathrm{t}=
L_n^\uparrow+L_n^\downarrow$ ($n=0,1,2$) of the transmission
coefficients are given by\cite{Cla96}

\begin{equation}
L_n^\sigma=\int_{-\infty}^{\infty}(E-E_\mathrm{F})^nT^\sigma(E)
\frac{\partial f(E,V,T)}{\partial E}{\mathrm d}E
\end{equation}

\noindent where $f$ is the Fermi distribution function, which depends on
voltage and temperature. Equation (\ref{Eq1}) can be expressed in
terms of measurable thermoelectric quantities: the electric
conductance ($G$), thermopower ($S$), Peltier coefficient ($\Pi$)
and the electronic contribution to the thermal conductance
($\kappa$):

\begin{equation}
\left(\begin{array}{c}\Delta V\\\dot Q\end{array}\right)=
\left(\begin{array}{cc}R&S\\\Pi&\kappa\end{array}\right)
\left(\begin{array}{c}I\\\Delta T\end{array}\right)
\end{equation}

\noindent where

\begin{equation}
G=\frac{e^2}{h}L_0^\mathrm{t}
\end{equation}

\begin{equation}
S=-\,\frac{1}{eT}\frac{L_1^\mathrm{t}}{L_0^\mathrm{t}}
\end{equation}

\begin{equation}
\Pi=\frac{1}{e}\frac{L_1^\mathrm{t}}{L_0^\mathrm{t}}
\end{equation}

\begin{equation}
\kappa=\frac{1}{hT}\left(L_2^\mathrm{t}-
\frac{L_1^\mathrm{t\,2}}{L_0^\mathrm{t}}\right)
\end{equation}

\noindent Notice that, according to these formulae, in order to produce
the highest thermopower it is necessary to have the factor in the
denominator ($L_0^\mathrm{t}=L_0^\uparrow+L_0^\downarrow$) as small as
possible. However, as we will see later, in a system with spin
polarization both channels are usually different around the Fermi level,
one of them being much larger than the other in some cases, which
decreases the value of $S$.

The figure of merit can also be expressed in terms of the transmission
moments by substituting the above expressions
in the $ZT$ definition:

\begin{equation}
ZT=\frac{1}{\frac{L_0^\mathrm{t}L_2^\mathrm{t}}
{L_1^\mathrm{t\,2}}-1}
\end{equation}

\noindent In this case the figure of merit becomes large when the
factor of the moments in the denominator decreases towards 1.

Approximations to these expressions can be obtained in the limit of
low temperatures by expanding $T(E)$ about $E=E_\mathrm{F}$, which
we take equal to 0 eV. In case of a single level coupled to featureless
leads, whose transmission is given by a Breit-Wigner resonance,

\begin{equation}
T(E)=\frac{\Gamma^2}{(E-\epsilon_0)^2+\Gamma^2}
\end{equation}

\noindent where $\Gamma$ is the strength of coupling of the level
to the leads and $\epsilon_0$ is the energy of the level, the low-T bias
conductance, Seebeck coefficient and figure of merit are given by:

\begin{eqnarray}
G&=&\mathrm{G}_0\,\frac{\Gamma^2}{\epsilon_0^2+\Gamma^2}\\
S&=&-\,\mathrm{G}_\mathrm{th}\,\frac{2\,h}{e}\,\frac{\epsilon_0}{\epsilon_0^2+
\Gamma^2}\\
ZT&=&\frac{4\,\mathrm{S}_0\,\epsilon_0^2}{(\epsilon_0^2+\Gamma^2)^2-
4\,\mathrm{S}_0\,\epsilon_0^2}
\end{eqnarray}

\noindent where $\mathrm{G}_0=2e^2/h$ and $\mathrm{G}_\mathrm{th}=
\pi^2k_\mathrm{B}^2T/3h$ are the electrical and thermal conductance
quantum units, and $S_0=h\,T\,\mathrm{G}_\mathrm{th}$.

For a Fano resonance produced by a side level with the same on-site
energy as the backbone level ($\epsilon_0$), and coupled to this
last one by a matrix element $V$, the transmission is

\begin{equation}
T(E)=\frac{\Gamma^2(E-\epsilon_0)^2}{\left[(E-\epsilon_0)^2-
V^2\right]^2+\Gamma^2(E-\epsilon_0)^2}
\end{equation}

\noindent and the thermoelectric coefficients $G$ and $S$ and
figure of merit are given by

\begin{eqnarray}
G&=&\mathrm{G}_0\,\frac{\Gamma^2\epsilon_0^2}{(\epsilon_0^2-V^2)^2+
\Gamma^2\epsilon_0^2}\\
S&=&-\,\mathrm{G}_\mathrm{th}\,\frac{2\,h}{e}\,\frac{\Delta(0)}
{\epsilon_0\Gamma^2}\\
ZT&=&\frac{\mathrm{S}_0\,\Delta^2(0)}{\epsilon_0^2\Gamma^4-
\mathrm{S}_0\,\Delta^2(0)}
\end{eqnarray}

\noindent where $\Delta(0)= \left[(2\,\epsilon_0^2- 2\,V^2+\Gamma^2)T(0)-
\Gamma^2\right]$. Notice the Fano resonance diverges at $\epsilon_0=0$, which
is an unphysical singularity. This can be avoided by including a background
transmission due to other resonances, which are always present in realistic
systems.

With these expressions it is then possible to calculate the
thermoelectric coefficients in a given junction, provided the
transmission is known. In general, in order to obtain large $S$
and $ZT$ it is convenient to have large derivatives around the
Fermi level, since the first moment ($L_1$) is, at least at low
temperatures, proportional to the derivative of the transmission.
Notice again that these expressions can only be used at low temperatures,
but qualitative trends derived from them (i.e. the overall shape
of the thermoelectric coefficients as a function of the level position)
are still valid at large temperatures.

\section{First principles calculations}\label{FirstP}

Molecular junctions that show sharp features around the Fermi level
could very good candidates to act as thermoelectric enhancers. In
particular, molecular junctions based on metalloporphyrin wires, which
have recently been subjected to a lot of interest, both
theoretically\cite{Rov97,Lia02,Pal09,Wan09,Gar13,Fer13} and
experimentally\cite{Ots10,Sed11}, show, for certain metallic elements,
a series of resonances close to the Fermi level which can be Breit-Wigner
or Fano-like. This property makes them specially appealing for
thermoelectricity, since such resonances can be employed to finely tune
the thermoelectric response with a gate voltage (see below).

In order to have a clear picture of the influence of the electronic
properties on the thermoelectric response we initially calculated from
first principles the electronic and transport properties of
metalloporphyrin dithiolate molecules between gold electrodes. In next
section we pay attention to the most important features of these systems
with the help of a simple model. The metallic elements which produce
states close to the Fermi level are Fe and Cu atoms\cite{Fer13}. We
therefore focus on this study only on Fe and Cu metalloporphyrins
dithiolate.

The first-principles calculations were performed in the framework of
density functional theory (DFT)\cite{Koh65}. We used the SIESTA
code\cite{SIESTA}, which employs norm conserving pseudopotentials and
a basis set of pseudoatomic orbitals. We included non-linear core
corrections\cite{Lou82} in the transition-metal pseudopotentials to
correctly account for the overlap between the valence and the core
states. We used for gold a single-${\zeta}$ basis (SZ) with explicit
$s$ and $d$ orbitals as valence orbitals. For all the other elements
(H, C, O, N, S and transition-metal) we used a double-${\zeta}$ polarized basis (DZP).
The exchange and correlation potential was approximated with the
generalized gradient approximation(GGA), as parameterized by Perdew,
Burke and Ernzernhof\cite{PBE}. We defined the real space grid with an
energy cutoff of 400 Ry. We performed the structural relaxations and
transport calculations in the $\Gamma$ point, which was enough to relax
the coordinates and correctly determine the transmission around the Fermi
level. We also did tests with $k-$points (2x2 in the perpendicular
directions) and the results were essentially the same around the Fermi
level. We relaxed the coordinates until all forces were smaller than
0.01 eV/\AA.

We corrected the self-interaction and other errors produced by DFT
by using the DFT$+U$ approach, which yields qualitatively correct
results in systems with transition-metal atoms\cite{Ani91}. Adding this
parameter to the central metallic atom was equivalent to adding a gate
potential which moves the states associated to it, as we shall see.
Take into account however that without a gate voltage only the results
with $U$ can be trusted. To reproduce previous theoretical results for
the gas-phase iron metalloporphyrin\cite{Men02,Pan08}, we used $U=4.5$ eV.
This value was also employed in the molecule between electrodes and in
other metallic elements. Notice that small differences in the $U$-term due
to the electrodes or other metallic atoms do not affect the results and the
only effect is a slight movement of the resonances around the Fermi
level, which produces qualitatively similar results.

Structurally, the gold electrodes were grown in the (001) direction.
The sulfur atoms were contacted to the gold surfaces in the hollow
position, which was found to be more stable than the top and bottom
configurations, at a distance of 1.8 \AA. The transport calculations
were performed with the Smeagol and Gollum codes\cite{Sme,Gol}. According
to the transport formalism junction was divided in three parts: left and
right leads and extended molecule (EM), which included the central part of
the junction and also some layers of the gold leads to make sure that the
electronic structure was converged to the bulk electronic structure away
from the surfaces.

\begin{figure}
\includegraphics[width=\columnwidth]{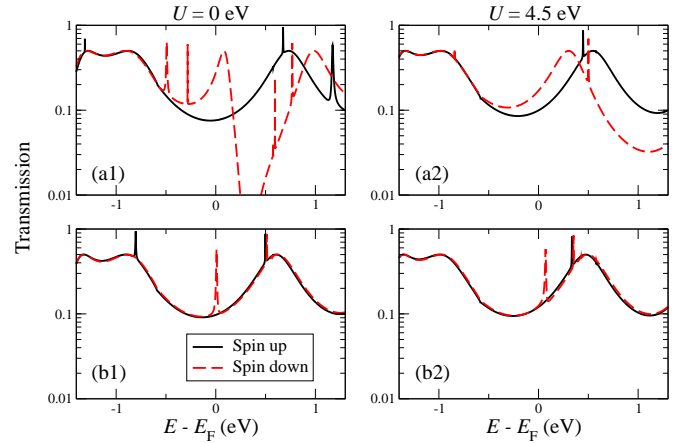}
\caption{\label{TRC.GGA}(Color online) Transmission coefficients
for Fe (a) and Cu (b) metalloporphyrins dithiolate between gold
electrodes, calculated with GGA and $U=0$ eV (1) and $U=4.5$ eV (2).}
\end{figure}

We show in Fig. (\ref{TRC.GGA}) the transmission of Fe and Cu
metalloporphyrins dithiolate between gold electrodes, with and without
the $U$ correction. As can be seen the bare Fe case shows a very clear
Fano resonance with its antiresonance close to the Fermi level. When
the $U$-term is added the resonance moves to higher energies and its effect
on the transmission around the Fermi level decreases. In the Cu case,
however, there seems to be a sharp Breit-Wigner resonance which moves
also to higher energies when the $U$-term is included.

\begin{figure}
\includegraphics[width=\columnwidth]{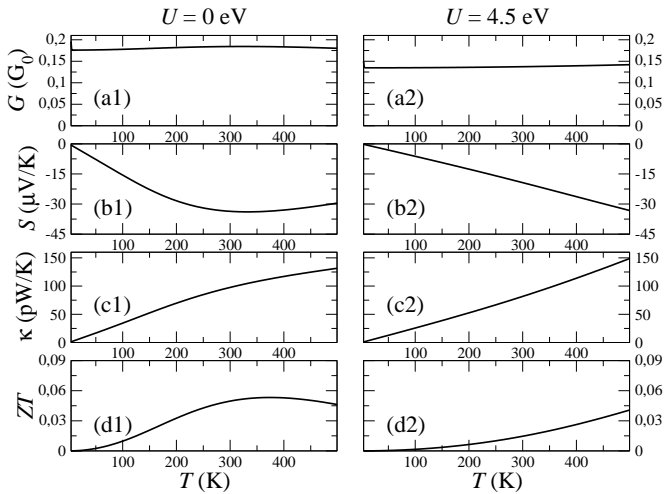}
\caption{\label{Thp.Fe.GGA}(Color online) Thermoelectric coefficients
for Fe metalloporphyrins dithiolate between gold electrodes, calculated
with GGA and $U=0$ eV (1) and $U=4.5$ eV (2). From top to bottom,
conductance $G$ (a), Seebeck coefficient $S$ (b), thermal conductance
$\kappa$ and figure of merit $ZT$.}
\end{figure}

\begin{figure}
\includegraphics[width=\columnwidth]{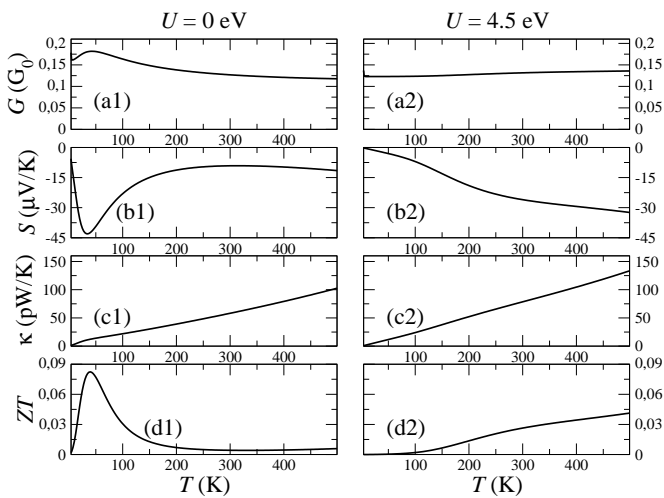}
\caption{\label{Thp.Cu.GGA}(Color online) Thermoelectric coefficients
for Cu metalloporphyrins dithiolate between gold electrodes, calculated
with GGA and $U=0$ eV (1) and $U=4.5$ eV (2). From top to bottom, conductance
$G$ (a), Seebeck coefficient $S$ (b), thermal conductance $\kappa$ and
figure of merit $ZT$.}
\end{figure}

From the transmissions we calculate the thermoelectric properties
by using the equations in section \ref{ThCoeff}. The results are
shown in Figs. (\ref{Thp.Fe.GGA}) and (\ref{Thp.Cu.GGA}). The
temperature dependence on the horizontal axis enters in the
Fermi distribution function, as explained before. In the case of
iron the electric conductance is almost constant, whereas the
thermal conductance increases roughly linearly with temperature.
The Seebeck coefficient, which is negative and signals that the
Fermi level is close to the LUMO, is relatively large and its
evolution with temperature qualitatively changes when the $U$-term is
included and the state moves to higher energies. The same
happens to the figure of merit, but it is rather small. These
evolutions can be explained by taking into account the spin-polarized
Fano resonance, which is a bit above the Fermi level and produces
large changes in the derivative of the transmission ($L_1$) without
$U$ but move to higher energies when the $U$-term is included and therefore
leave an smoother transmission at the Fermi level.

\begin{figure}
\includegraphics[width=\columnwidth]{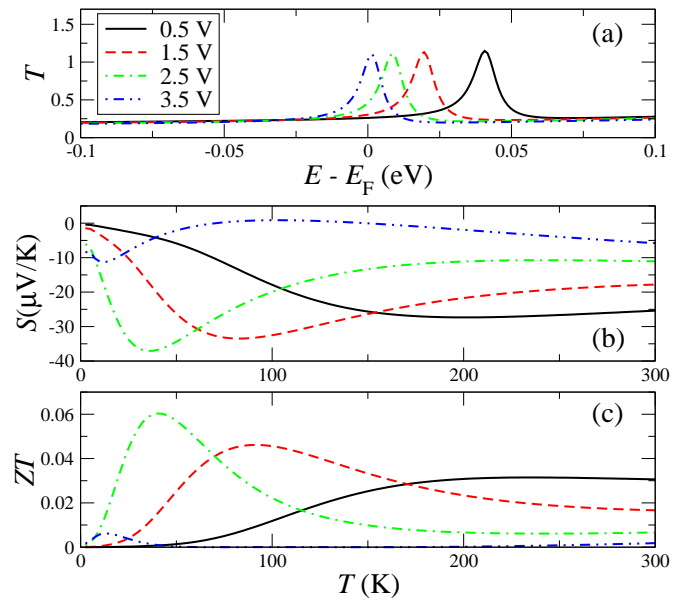}
\caption{\label{Thp.Gate.GGA}(Color online) Spin-polarized transmission as
a function of energy (a) and Seebeck coefficient (b) and figure of merit
(c) as a function of temperature for Cu metalloporphyrins dithiolate between
gold electrodes, calculated with GGA, $U=4.5$ eV and different gate
potentials.}
\end{figure}

\begin{figure}
\includegraphics[width=\columnwidth]{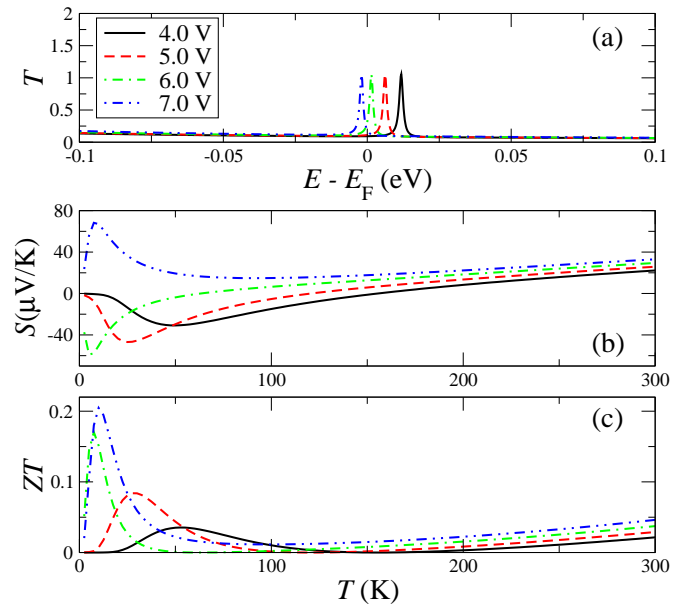}
\caption{\label{Thp.Gate.d_0.8.GGA}(Color online) Spin-polarized transmission
as a function of energy (a) and Seebeck coefficient (b) and figure of merit
(c) as a function of temperature for stretched Cu metalloporphyrins dithiolate
between gold electrodes (the contact distance between the molecule and the
electrodes was increased 0.8 \AA on each side from the equilibrium
configuration), calculated with GGA, $U=4.5$ eV and different gate potentials.}
\end{figure}

The copper case is more interesting, as the temperature evolution
of some quantities has more features and changes more dramatically
when the state moves. Again, the electric conductance is rather
constant and the thermal conductance increases linearly. The
Seebeck coefficient shows however a dip at low temperatures, and
becomes almost constant as the temperature increases. The dip
disappears when the state moves to higher energies and the
magnitude decreases to more negative values with $T$. The figure
of merit is again small but has a peak at low temperatures. Such
peak disappears when the $U$-term is included and is substituted by
a smooth increase. This evolution is a consequence of the presence
at the Fermi level of a sharp spin-polarized resonance which moves
closer to the LUMO when the $U$-term is included. Such resonance, which is
just a bit above the Fermi level gives rise to large derivatives and
therefore dramatically increases, in absolute value, the Seebeck
coefficient $S$. This coefficient is however not very large because
the sum of the transmission ($L_0$) of both spin channels is not
small.

The copper molecule could be a candidate to show large thermoelectric
properties due to the presence of a resonance close to the Fermi level.
However, the introduction of the $U$ moves the resonance to higher
energies and decreases the thermoelectric response. It is then interesting
to consider the case of applying a gate voltage that brings back the
resonance to the Fermi level. We show that in Fig. (\ref{Thp.Gate.GGA}).
As can be seen, moving the resonance to lower energies increases the
absolute value of  both $S$ and $ZT$\cite{Phk}. The highest values are
obtained when the largest slope of the resonance is just at the Fermi
level ($V_\mathrm{G}=2.5$ V). Beyond that point $S$ and $ZT$ decrease
($V_\mathrm{G}=3.5$ V). The increase of the thermoelectric properties
is not very spectacular however due to the fact that the resonance is
not extremely sharp. More acute resonances can be however obtained by
decreasing the coupling between the molecule and the gold electrodes,
which reduces the width of all transmission features. We show in Fig.
(\ref{Thp.Gate.d_0.8.GGA}) results for a system where the distance
between the gold electrodes and the sulphur atoms on each side
increases by 0.8 \AA\ relative to the equilibrium configuration. As
can be seen, now the resonance is very sharp and the changes in the
thermoelectric properties are more spectacular. Notice that the
thermopower changes sign when the resonance crosses the Fermi level
due to the change of slope. Based on these results we can claim that
stretching molecular electronic junctions increase in general the
thermoelectric performance.

\section{Tight-binding model}\label{TightB}

\subsection{General properties}

The most important features in the transmission of metalloporphins
dithiolate junctions can be reproduced with a simple model. With such
model we have analysed in detail the impact of Fano resonances on the
charge transport properties of these systems\cite{Fer13}. We have found
that these molecules contribute with three broad resonances to the
transmission coefficients of the junctions, which correspond to the
HOMO-1 (a $\sigma$ molecular orbital), the HOMO and the LUMO
(that has $\pi$ character). In addition, the 3d-atom contributes with
a spin-polarized strongly localized state, which hybridizes with the
HOMO-1 or the LUMO, depending on the element, and gives rise to a Fano
resonance. To model these junctions, we use the Hamiltonian

\begin{eqnarray}
\hat {\cal H}&=&\hat{\cal H}_\mathrm{gold}+\hat{\cal H}_\mathrm{M}+
\hat{\cal H}_\mathrm{gold-M}\\
\hat{\cal H}_\mathrm{gold}&=&\sum_{k,\sigma} \,\epsilon_k \,c_{k\sigma}^\dagger\,
c_{k\sigma}\\
\hat{\cal H}_\mathrm{M}&=&\sum_\sigma\,\epsilon_{d\sigma}\,\hat d_\sigma^\dagger\hat
d_\sigma\,+\,\sum_{i=1,2,3,\sigma}\,\epsilon_i\,\hat c_{i\sigma}^\dagger
\hat c_{i\sigma}+\nonumber\\
&&+\sum_{i=1,3} V_{i}\,\left(
\hat c_{i\sigma}^\dag\hat d_\sigma+\hat d_\sigma^\dag\hat
c_{i\sigma}\right) \\
\hat{\cal H}_\mathrm{gold-M}&=&\sum_{k,i,\sigma} T_{i} \,(c_{k\sigma}^\dagger\,c_{i\sigma}+c_{i\sigma}^\dagger\,c_{k\sigma})
\end{eqnarray}

\begin{figure}
\includegraphics[width=\columnwidth]{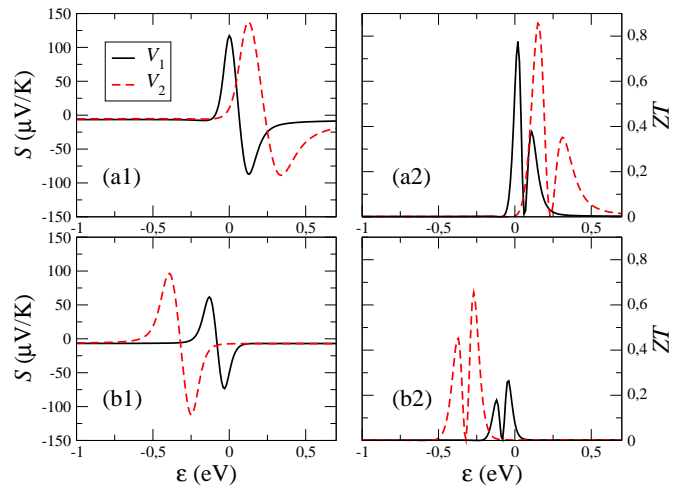}
\caption{\label{Thp.Model} Seebeck coefficient $S$ (1) and figure of
merit $ZT$ (2) for a model system with a Fano (a) and a
Breit-Wigner-like (b) resonance in the HOMO-LUMO gap, calculated
at $T=250$ K. In the case of the Fano (Breit-Wigner-like) resonance the
coupling between the $\pi$ ($\sigma$) level and the $d$ level is
$V_1=0.2$ eV ($V_2=0.4$ eV), which corresponds to the continuous (dashed)
lines.}
\end{figure}

\noindent where the operators $\hat c_{i,\sigma},\,i=1,2,3$ represent
the HOMO-1, HOMO and LUMO molecular levels. The operator $\hat d$
represents the $d$ level associated to the central metallic atom.
We assume that this level couples only to either the HOMO-1 or to the LUMO
levels (e.g.: only $V_1$ or $V_3$ are different from zero). We assume a wide
band approximation for the band structure of the gold electrodes, so that
its density of states $\rho_e$ and therefore the Gamma matrices
($\Gamma_{i}=T_{i}^2\,\rho_e$) are constant. Finally, we assume that the
$d$-level is spin-polarized so that only the spin-up $d$-level enters the
relevant energy window.

By using this model we have found\cite{Fer13} that the presence of
two types of resonances can be explained by how the $d$ state couples
to other molecular states. If the $d$ state couples to the HOMO-1, which
is a $\sigma$-like molecular orbital, the state produces around the
Fermi level a resonance which looks like a Breit-Wigner resonance. Such
peak comes really from a Fano resonance, whose dip is note seen because
it is masked by the larger transmission around the Fermi level. If the $d$
state couples however to the LUMO orbital a clear Fano resonance appears
around the Fermi level because the dip affects the transmission of the
LUMO, which is not completely masked by the transmission of other states.

\subsection{Thermoelectric properties}

From the transmission it is easy to obtain the thermoelectric
coefficients by using the equations in section \ref{ThCoeff}. We focus
specially in the thermopower and figure of merit, which are the most
relevant for thermoelectric efficiency. Since we can easily vary the
parameters of the model we study different effects such as the movement
of the states across the Fermi level and the change of the coupling
between the $d$ state and other molecular states or between the
molecule and the leads. We show in Fig. (\ref{Thp.Model}) the
thermopower and figure of merit calculated at $T=250$ K for Fano
and Breit-Wigner-like resonances as a function of the position
of the state that gives rise to them. As can be seen the thermopower
shows a peak-dip structure, which is a consequence of the change
of the derivative of the transmission as the resonance crosses
$E_\mathrm{F}$. Such structure is asymmetric in the first case
because the Fano resonance is also asymmetric, as it is made of
a resonance followed by an antiresonance. In the second case it
is also slightly asymmetric because the resonance is not exactly
in the middle of the HOMO-LUMO gap. The figure of merit, which
reaches values as high as 0.86, follows roughly the square of the
thermopower and therefore has two peaks and a dip, which corresponds
to the highest point of each resonance. Notice also both quantities
$S$ and $ZT$ are larger in the Fano case due to more pronounced changes
in the derivative when the antiresonance follows the resonance.

\begin{figure}
\includegraphics[width=\columnwidth]{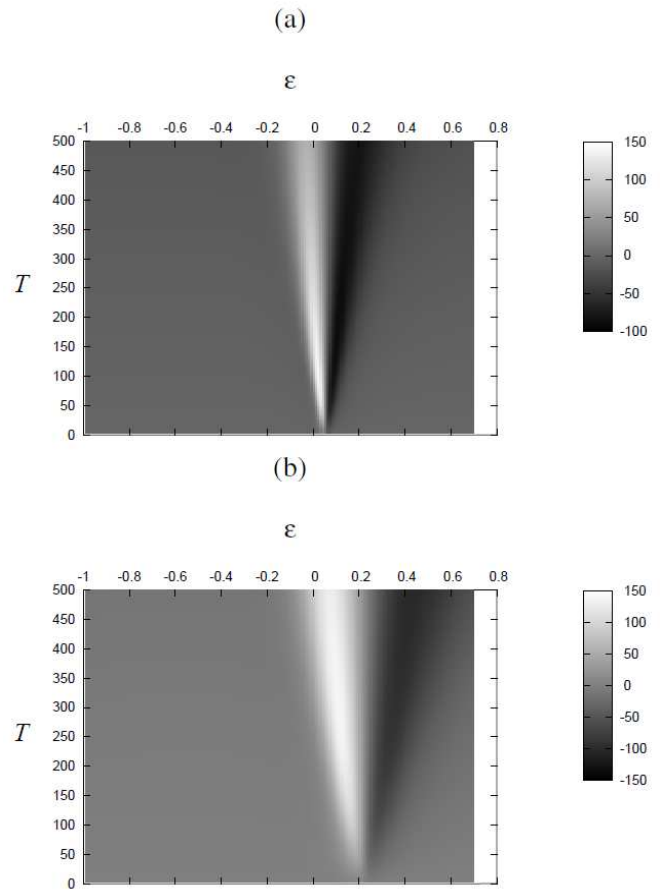}
\caption{\label{S.FR} Seebeck coefficient $S$ in units of $\mu$V/K
for model metalloporphyrins dithiolate which have a Fano-like resonance
around the Fermi level, as a function of temperature $T$ (in Kelvin)
and the level position $\epsilon_0$ (in eV). The coupling between the
$d$ state and the $\pi$ state is 0.2 eV (a) and 0.4 eV (b).}
\end{figure}

\begin{figure}
\includegraphics[width=\columnwidth]{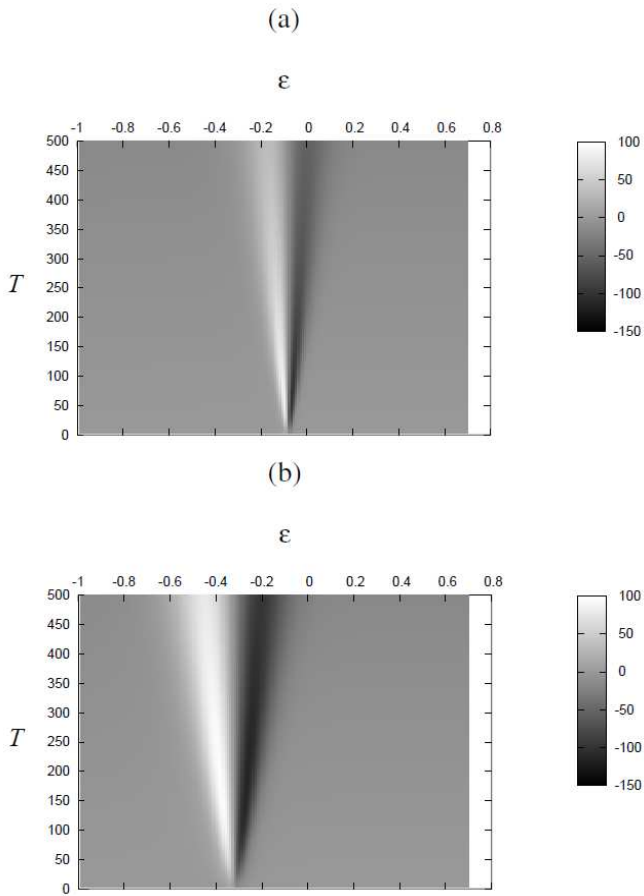}
\caption{\label{S.BW} Seebeck coefficient $S$ in units of $\mu$V/K
for model metalloporphyrins dithiolate which have a Breit-Wigner-like
resonance around the Fermi level, as a function of temperature $T$
(in Kelvin) and the level position $\epsilon_0$ (in eV). The coupling
between the $d$ state and the $\sigma$ state is 0.4 eV (a) and 0.8 eV
(b).}
\end{figure}

One effect that can influence the values of the thermopower and
figure of merit is the coupling $V$ between the $d$ level and
the corresponding molecular level. This can be done e.g. by using
a different metallic atom or straining/compressing the molecule.
As can be seen, changing such coupling increases the absolute value
of both quantities, specially in the Breit-Wigner-like case, but the
change is not very large because the only difference in the
transmission is due to an increase of the width of the resonances,
which does not affect the derivative too much. This seems to indicate
a relative robustness of the absolute value of $S$ and $ZT$ as
a function of the metallic atom or small molecular conformation
changes. When the coupling increase there is also a movement of the
peaks and dips to lower or higher energies, which is produced by the
increase of the separation between levels.

\begin{figure}
\includegraphics[width=\columnwidth]{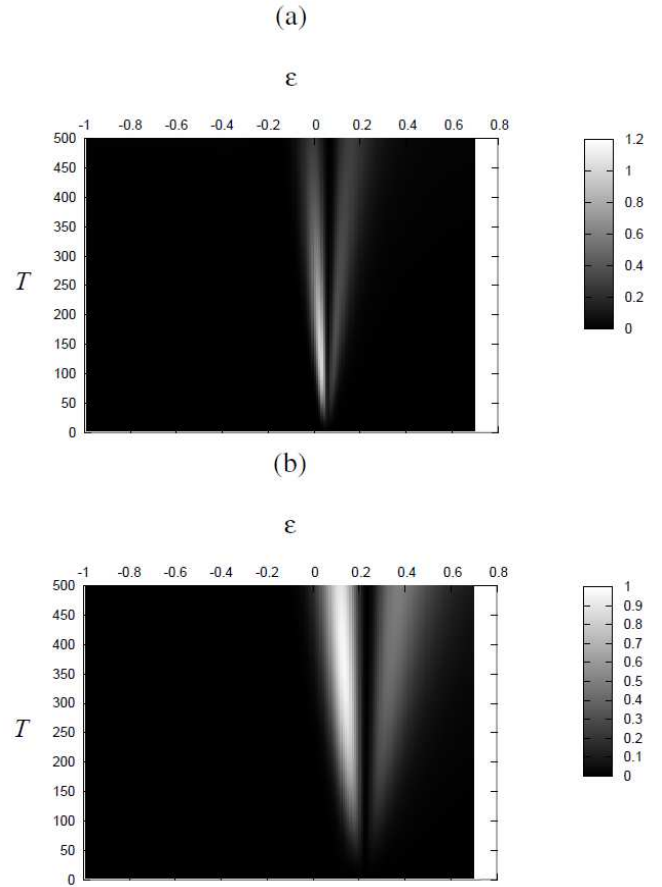}
\caption{\label{Z.FR} Figure of merit $ZT$ for model metalloporphyrins
dithiolate which have a Fano-like resonance around the Fermi level, as
a function of temperature $T$ (in Kelvin) and the level position
$\epsilon_0$ (in eV). The coupling between the $d$ state and the
$\pi$ state is 0.2 eV (a) and 0.4 eV (b).}
\end{figure}

\begin{figure}
\includegraphics[width=\columnwidth]{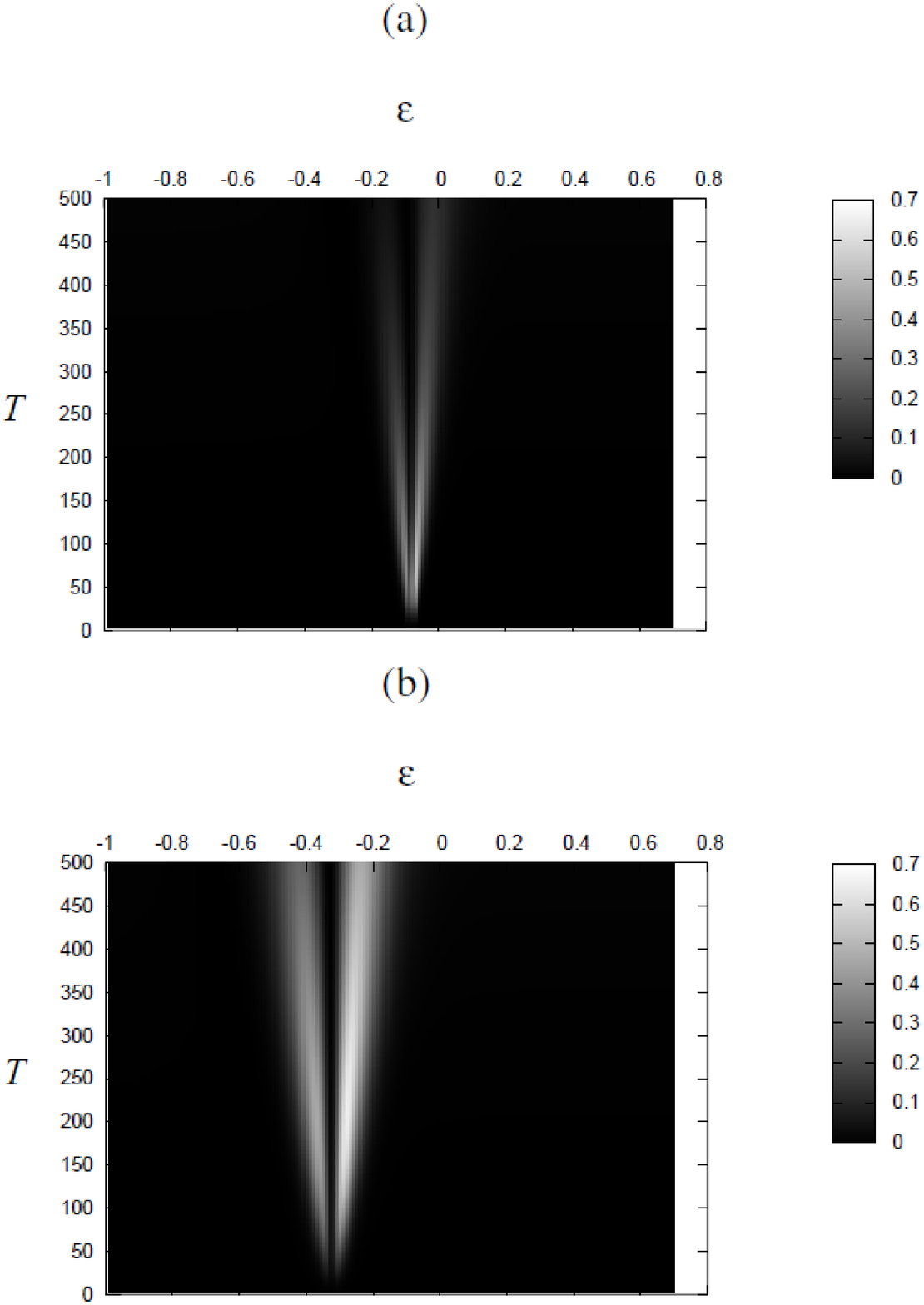}
\caption{\label{Z.BW} Figure of merit $ZT$ for model metalloporphyrins
dithiolate which have a Breit-Wigner-like resonance around the Fermi
level, as a function of temperature $T$ (in Kelvin) and the level
position $\epsilon_0$ (in eV). The coupling between the $d$ state
and the $\sigma$ state is 0.4 eV (a) and 0.8 eV (b).}
\end{figure}

The total evolution of both quantities as a function of the level
position and temperature, for both couplings, is shown in Figs.
(\ref{S.FR}), (\ref{S.BW}), (\ref{Z.FR}) and (\ref{Z.BW}). As can
be seen the peaks-dips structures remain the same for large
temperature ranges. Both quantities show high peaks and dips at
relatively low and intermediate temperatures, up to $\sim 300$ K,
and tend to slightly decrease beyond room temperature. According
to this, the most efficient heat to electricity conversion (large
$S$ and $ZT$) can be achieved in the case Fano resonances at
temperatures close to room temperature.

\begin{figure}
\includegraphics[width=\columnwidth]{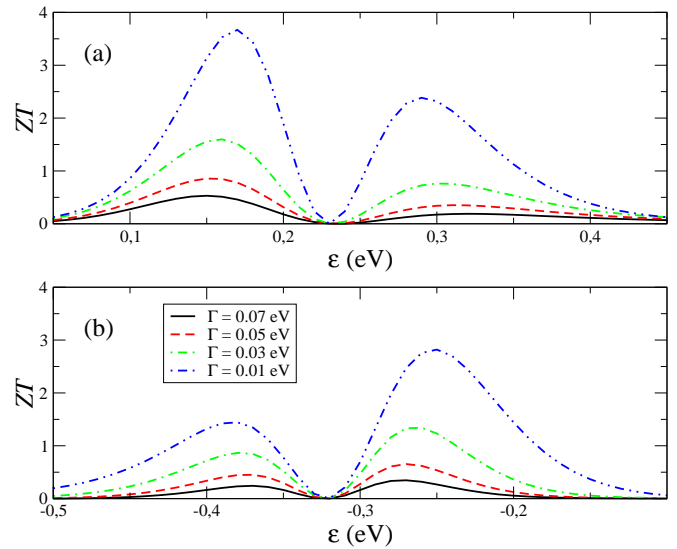}
\caption{\label{ZT.Model} Figure of merit $ZT$ for a model
metalloporphyrin dithiolate with a Fano (a) and a
Breit-Wigner-like (b) resonance in the HOMO-LUMO gap, calculated
at $T=250$ K, for various coupling strengths between the HOMO and
LUMO and the electrodes. The coupling between the $d$ state and the
$\pi$ or $\sigma$ states is (a) 0.4 and (b) 0.8 eV.}
\end{figure}

The figure of merit is large but not too much. As previously stated,
the value of the figure of merit is capped due to the relatively large
transmission at the Fermi level and the presence of the other spin
channel. The transmission of both channels can however be decreased
by reducing the coupling between the molecule and the electrodes,
which decreases the width of all transmission resonances and reduces
the transmission in the middle. We show in Fig. (\ref{ZT.Model})
the figure of merit calculated around the peaks-dip structure for
various coupling strengths or $\Gamma$ matrices\cite{Coupling} between
the levels and the electrodes. As can be seen, the smaller the coupling
or the corresponding $\Gamma$ matrix, the larger the figure of merit.
For small couplings it can reach values as large 3.7. For such small
couplings a note of caution should be added however since the width
of the resonances is so small that strong correlations could change
the picture of the physical properties.

\section{Conclusions}

The thermoelectric properties of junctions with states close
to the Fermi level have been calculated using analytical derivations
and a simple model. A spin-polarized first-principles calculation of
a junction made of a metalloporphyrin dithiolate molecule between gold
electrodes has also been included as a realistic example. The Fano and
Breit-Wigner-like resonances greatly enhance the thermopower and figure
of merit when they cross the Fermi level. The maximum value of these
quantities depends on the coupling between the state that gives rise
to the resonance and the other molecular states. The bigger the coupling
the bigger the thermopower and figure of merit. Their evolution
with temperature has also been studied and it was found that the
largest efficiency, corresponding to the largest figure of merit,
can be achieved at temperatures close to room temperature. Finally,
the coupling between the molecule and the electrodes was also taken
into account and it was found that reducing it greatly enhances the
figure of merit, which can reach values larger than 1.

The research presented here was funded by the Spanish MICINN through
the grant FIS2012-34858 and by the Marie Curie network NanoCTM.
VMGS thanks the Spanish Ministerio de Econom\'{\i}a y Competitividad
for a Ram\'on y Cajal fellowship (RYC-2010-06053). RRF thanks
Consejer\'{\i}a de Educaci\'on del Principado de Asturias for a
Severo Ochoa grant (BP11-069).


\begin{thebibliography}{References}
\bibitem{SpinPol}
Notice that in general, in cases with spin polarization and no
spin-orbit coupling, the orientation between the spin and the
lattice coordinates is arbitrary and therefore it is not possible
to determine the spin-filtering behavior of the system without
coupling to a ferromagnetic material.
\bibitem{Phk}
Notice the value of the figure of merit would be further decreased by
taking into account the phonon thermal conductance, $\kappa_p$, i.e.
$ZT=S^2GT/(\kappa+\kappa_p)$. In this article we consider only the
electronic contribution. In general, the phonon contribution, which
can be obtained by calculating the dynamical matrix\cite{Mar09}, can
be relatively large at high temperatures if perfect electrodes are used.
However, it can be greatly decreased by nanostructuring the electrodes to
include phonon band gaps or using disordered electrodes that increase
phonon scattering\cite{Fag99,Kam99,Dub11}.
\bibitem{Coupling}
As a simplification we assume the coupling of the HOMO-1 does not change.
This is in general not correct since the coupling of all states changes
when the molecule gradually uncouples from the leads, but will allow us
to see just the effect of decreasing the coupling of the HOMO and LUMO.
Notice decreasing the HOMO-1 coupling would further enhance the figure of
merit, sine the transmission would be even lower at the Fermi level.
This is also the reason why the increase of the figure of merit saturates
for very small couplings.
\end{thebibliography}
\end{document}